\documentclass[aip,apl,reprint,twocolumn,a4paper,showkeys,showpacs,balancelastpage]{revtex4-1}

\usepackage[british]{babel}
\usepackage{graphicx}
\usepackage{natbib}
\usepackage{amsmath}

\begin{document}

\date{\today}

\title{Formation and control of wrinkles in graphene by the wedging transfer method}

\author{V.E. Calado}
\email[]{v.e.calado@tudelft.nl}

\author{G.F. Schneider}

\author{A.M.M.G. Theulings}

\author{C. Dekker}

\author{L.M.K. Vandersypen}
\affiliation{Kavli Institute of Nanoscience, Delft University of Technology, 2628 CJ Delft, The Netherlands}

\begin{abstract}
We study the formation of wrinkles in graphene upon wet transfer onto a target substrate, whereby draining of water appears to play an important role. We are able to control the orientation of the wrinkles by tuning the surface morphology. Wrinkles are absent in flakes transferred to strongly hydrophobic substrates, a further indication of the role of the interaction of water with the substrate in wrinkle formation. The electrical and structural integrity of the graphene is not affected by the wrinkles, as inferred from Raman measurements and electrical conductivity measurements.
\end{abstract}
\pacs{72.80.Vp}

\maketitle

Graphene has received a lot of attention since its isolation from graphite.\cite{Novoselov2004} 
Although graphene is a 2D crystal, it is so far only found on a substrate, as a membrane with a supporting construction\cite{Meyer2007} or grown at the surface of SiC.\cite{deHeer2007} Therefore it is considered to be a quasi 2D crystal. Quasi-2D graphene is in general not flat, but has a tendency to from corrugations, including ripples, wrinkles and bubbles.\cite{Fasolino2007,Meyer2007}
The curvature associated with such corrugations is predicted to alter graphene's electronic and structural properties.\cite{Guinea2008may,Pereira2010} Corrugations are often regarded as undesirable, but they can be exploited for inducing pseudo-magnetic fields,\cite{Levy2010} creating chemically reactive sites,\cite{Balog2010} and for specific device applications such as optical lenses.\cite{Georgiou2011} Numerical simulations have been done to understand the formation of wrinkles and their impact on graphene.\cite{Wang2009,Gil2010,Duan2011,Min2011}

Wrinkles are commonly found in chemical-vapor-deposition (CVD) grown graphene that is transferred to other substrates.\cite{Liu2011,Ni2012,Gao2012,Zhu2012} In CVD graphene wrinkles  are formed at metal step edges due to thermal stress. The morphology of the metal growth surface can still be seen after transfer. In many cases additional wrinkles, ripples, and bubbles are formed upon transfer. Exfoliated graphene mostly conforms to the corrugations of the underlying substrate,\cite{Ishigami2007} although small additional wrinkles can be observed in a scanning tunnelling microscope.\cite{Xu2009}
 
 \begin{figure}[t!]
 \includegraphics{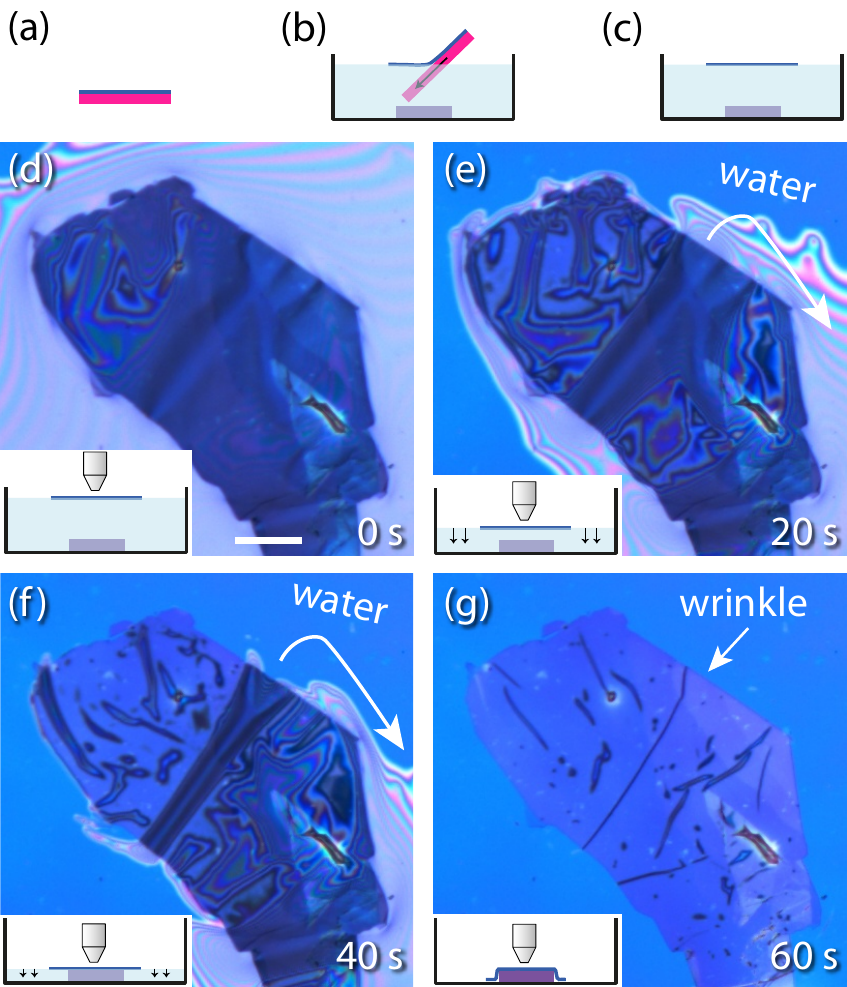}%
 \caption{(a) a SiO$_{2}$ substrate with graphitic flakes covered with a polymer. (b-c) The polymer film is ``wedged off'' the substrate and floats on water. (d-f) Filmstrip of a wet transfer process seen through an optical microscope (the time stamps are approximate). The arrow points at a large corrugation that is already present when the flake is floating on water. Once water is drained, the corrugation is left behind as a wrinkle. The scale-bar is $50~\mu$m. The schematics in the insets show the water level. \label{Wrinkle:fig0}}%
 \end{figure}
 
In this work we study the formation of wrinkles in graphene during the so-called wedging transfer process, a water-based transfer process.\cite{Schneider2010a}. We give insight in the driving forces for wrinkle formation and provide different routes to control the wrinkle orientation and abundance, or to eliminate wrinkles altogether. We also examine to what extent the electronic and structural integrity of graphene is preserved upon wrinkle formation.

The wrinkle formation can be easily seen with an optical microscope, as illustrated for a graphitic flake in Fig. \ref{Wrinkle:fig0}. Graphitic flakes of varying thicknesses are prepared by mechanical exfoliation of graphite (NGS Naturgraphit GmbH) on Si substrates with a $285$~nm thermal oxide. A hydrophobic polymer film covers the substrate (Fig. \ref{Wrinkle:fig0}(a)), and is ``wedged off'' the substrate by intercalation of water (Fig. \ref{Wrinkle:fig0}(b)), along with the graphitic flakes. As a result, the film with the flakes is floating on top of the water surface (Fig. \ref{Wrinkle:fig0}(c)). Then the water is gradually drained, which brings the polymer film closer to the target substrate underneath (Fig. \ref{Wrinkle:fig0}(d-e)). Here the target substrate is just another Si/SiO$_{2}$ substrate, but the transfer can be done onto arbitrary structures.\cite{Schneider2010a} As water is drained further the polymer film comes in contact with the target substrate and pre-existing wrinkles will act as channels that drain water (Fig. \ref{Wrinkle:fig0}(f)). Finally, the water is drained away completely and the channels are left behind as wrinkles (Fig. \ref{Wrinkle:fig0}(g)). 

\begin{figure}
 \includegraphics{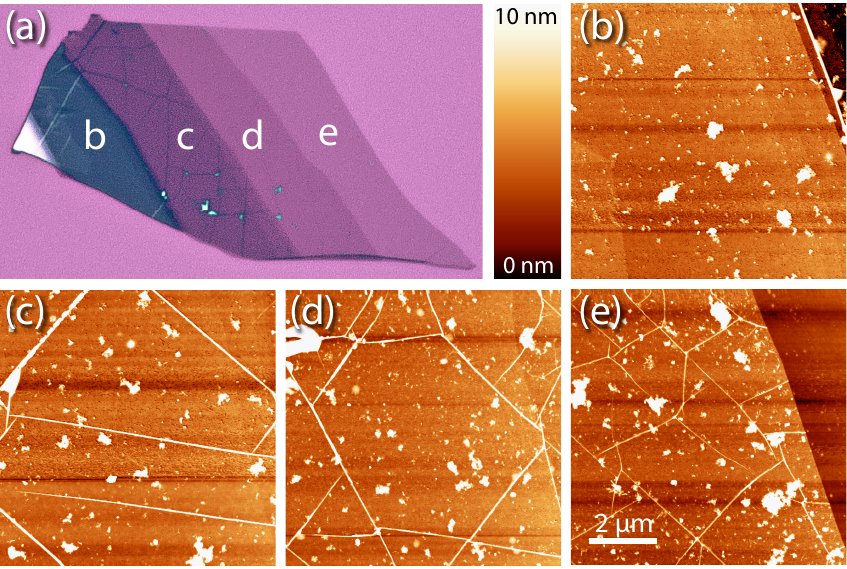}%
 \caption{(a) An optical microscope image of a wrinkled graphite flake of different thicknesses (down to ~3 nm thickness). (b-e) Atomic force microscope (AFM) image of the different parts as indicated in (a). \label{Wrinkle:Graphite}}%
 \end{figure}

As the thickness of a graphitic sheet is reduced (Fig. \ref{Wrinkle:Graphite}), the wrinkles increase in density and decrease in height (Fig. \ref{Wrinkle:Graphite}(b-e)). This can be expected given the lower stiffness of thinner sheets. This is seen in AFM images on different parts of the graphite flake ( Fig. \ref{Wrinkle:Graphite}(b-e)).
 
\begin{figure}
 \includegraphics{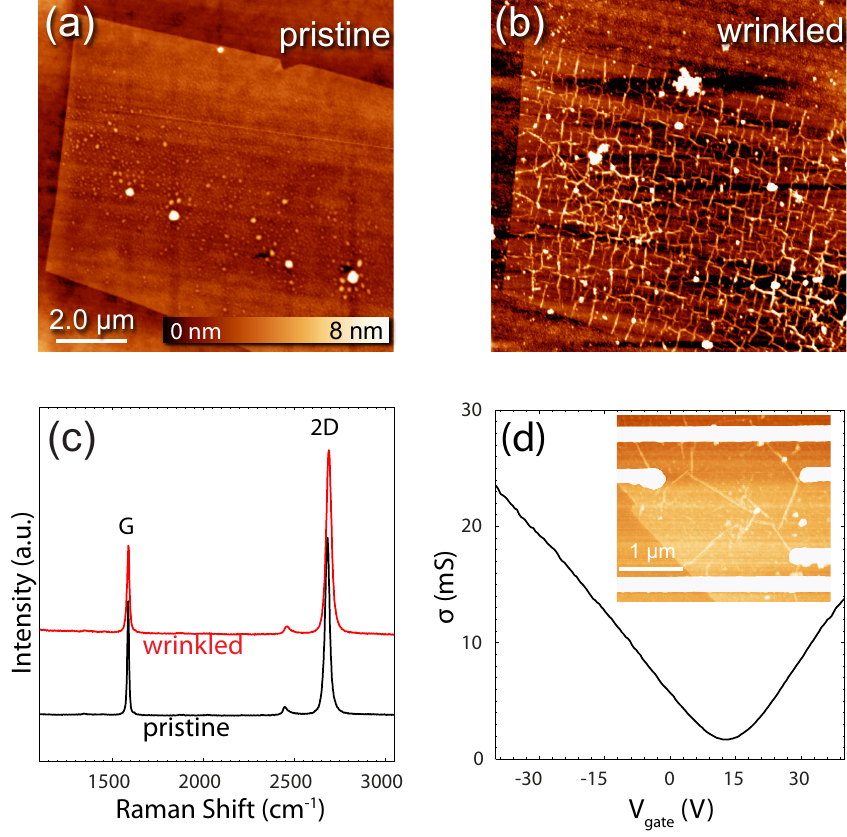}%
 \caption{(a) AFM image of a pristine SLG sheet as prepared by exfoliation. (b) An AFM image of the SLG sheet after transfer, now containing many wrinkles. (c) The Raman spectra of the pristine and wrinkled graphene layer of (a,b). (taken on a Renishaw Raman system 2000 with a $514$~nm argon laser, $1\mu$m spot size, $1$~mW power and spectral resolution of $\sim3$~cm$^{-1}$). (d) Conductivity as a function of applied back-gate voltage, measured at room temperature and at low pressure ($\sim10^{-5}$~mbar). Inset: AFM image of the device. \label{Wrinkle:fig1}}%
 \end{figure}

An AFM image of a single-layer graphene (SLG) flake, identified by its distinct Raman spectrum,\cite{Ferrari2006} is shown in Fig. \ref{Wrinkle:fig1}(a) before and \ref{Wrinkle:fig1}(b) after transfer.
Raman spectra on this flake taken before (pristine) and after (wrinkled) transfer are shown in Fig. \ref{Wrinkle:fig1}(c) The Raman spectra of pristine and wrinkled SLG are very similar, with no defect-related D band near $1350$~cm$^{-1}$ detectable within the noise level.

Next we perform electrical transport measurements in order to estimate the charge-carrier mobility and the mean-free path. We contact several wrinkled SLG flakes with $10/60$~nm Cr/Au electrodes using e-beam lithography. A typical device is shown in the inset of Fig. \ref{Wrinkle:fig1}(d). In a four-point field effect transistor (FET) geometry we measure the resistance as a function of applied back-gate voltage, Fig. \ref{Wrinkle:fig1}(d). The measurement is done in high vacuum ($\sim10^{-5}$~mbar) at room temperature. The flake geometries are irregular. For estimating the conductivity from the conductance we choose an aspect ratio such that the mobility values obtained are an underestimate. The charge-carrier density is calculated by applying the parallel plate capacitor model, taking the usual conversion factor for $285$~nm SiO$_{2}$ of $7.56\cdot10^{10}$~cm$^{2}$~V$^{-1}$. From the slope in Fig. \ref{Wrinkle:fig1}(d) we calculate a hole mobility of $3800$~cm$^{2}$~V$^{-1}$s$^{-1}$ and an electron mobility of $4400$~cm$^{2}$~V$^{-1}$~s$^{-1}$. These are typical values found for SLG on SiO$_{2}$. We have no indication that the wrinkles degraded the mobility. This is not surprising given that the mean free path is around $\sim20$~nm and the wrinkle-to-wrinkle distance is $\sim1\mu$m.

\begin{figure}
 \includegraphics{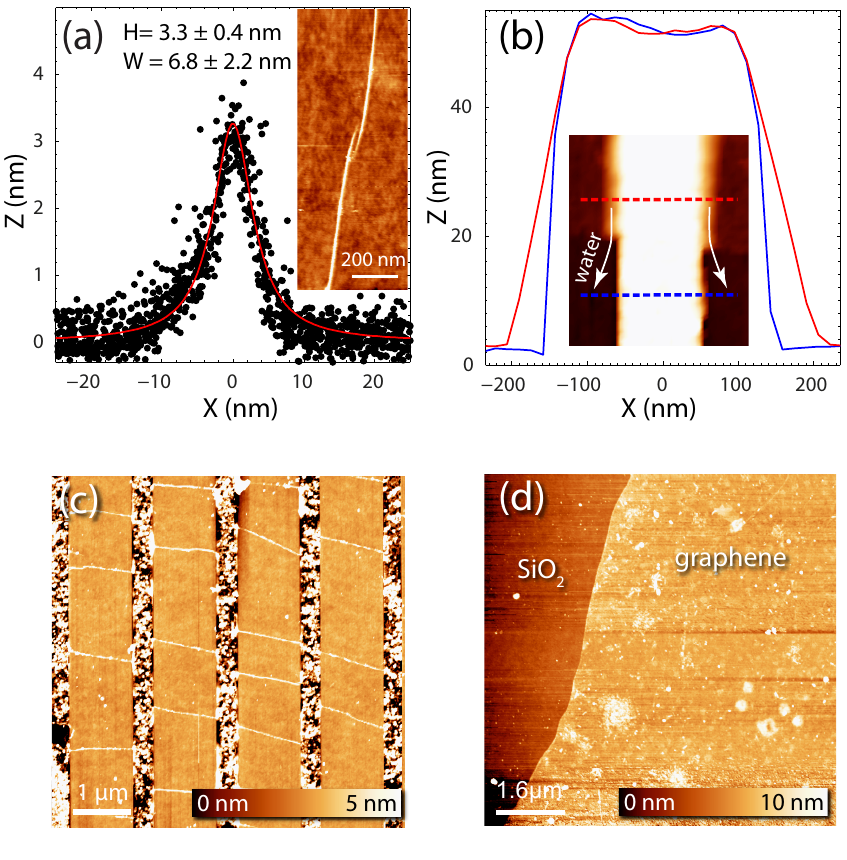}%
 \caption{(a) The height profile of a typical wrinkle. The black dots are a subset of data taken from the inset and plotted together on the same horizontal axis. The red curve is a fit, which is the result of an average over 294 traces. Inset: AFM image of a single wrinkle. The middle part is excluded in the fit. (b) The cross-section height profile across a bare metal strip on the substrate (blue) and across a strip covered with graphene (red), as indicated in the AFM image in the inset. (c) AFM image of SLG transferred onto a periodic step structure. The wrinkles tend to align perpendicularly to the steps. (d) AFM image of SLG transferred to a hydrophobic substrate. \label{Wrinkle:fig2}}%
 \end{figure}

Fig. \ref{Wrinkle:fig2}(a) shows an AFM height profile of a typical wrinkle. The mean height along the wrinkle is $H=3.3\pm0.4$~nm  with a mean FWHM of $W<6.8\pm2.2$~nm (due to the width of the AFM tip, the width measurement is an upper bound). These dimensions are well in the regime of $H^2/Wa>1$ ($a=0.24$ nm is the graphene lattice constant), where the pseudo-magnetic field from the corrugations is predicted to be large enough to induce a zero-energy Landau level.\cite{Guinea2008feb}

In Fig. \ref{Wrinkle:fig2}(b) we demonstrate that using the morphology of the substrate we are able to control the orientation of the wrinkles to a certain degree. We have fabricated 1~$\mu$m spaced metal strips of 200 nm wide and 55 nm high. SLG is transferred onto these periodic structures, and is partially suspended between the top edge of the step edges and the substrate, as can be seen in Fig. \ref{Wrinkle:fig2}(b). It shows that there is room for water to evacuate near the steps. Wrinkles form in the direction perpendicular to the steps (Fig. \ref{Wrinkle:fig2}(c)), presumably so that water can be evacuated towards the steps, and outwards from there. In this way we demonstrate the possibility to align nm sized wrinkles in graphene. We propose that milling small trenches ($<50$~nm width) in a substrate would yield a similar result, if graphene can freely suspend over these trenches. This could be used to study electronic transport across multiple aligned wrinkles.

Finally, we show that wrinkle formation can be prevented altogether by transferring flakes onto hydrophobic substrates, as shown in Fig. \ref{Wrinkle:fig2}(d) for a SLG flake. To render the surface hydrophobic, it was functionalized with a fluoroalkane molecule (CF$_{3}$(CF$_{2}$)$_8$(CH$_{2}$)$_2$SiCl$_3$, Sigma Aldrich). This observation supports the hypothesis that water plays a key role in wrinkle formation during wedging transfer.

In summary, we find that wet transfer of graphene onto commonly used hydrophilic substrates induces wrinkles in graphene. Raman measurements show no detectable defect-related D-peak in this wrinkled graphene, and transport measurements show mobilities comparable to those found before transfer. It is possible to control to a certain extent the density, size and orientation of the wrinkles. This could be useful for studying electronic transport across a controlled number of (aligned) wrinkles. It also opens up the possiblilty to explore a number of other experiments. In the presence of a magnetic field, the angle between field and flake will vary on a nm length scale. In addition, the curvature induces local pseudo-magnetic fields.\cite{Levy2010} The curvature also increases chemical reactivity which can be exploited to chemically bind atoms or molecules\cite{Elias2009,Nair2010,Balog2010} in one-dimensional patterns along the wrinkles.

We thank Gilles Buchs and Bart van der Linden for assistance with the Raman measurements, and Andrea Ferrari, Amelia Barrero, Xing Lan Liu and Stijn Goossens for discussions. This work is part of the research program of the Foundation for Fundamental Research on Matter (FOM), which is part of the Netherlands Organization for Scientific Research (NWO).

\bibliography{ref}

\begin{thebibliography}{24}%
\makeatletter
\providecommand \@ifxundefined [1]{%
 \@ifx{#1\undefined}
}%
\providecommand \@ifnum [1]{%
 \ifnum #1\expandafter \@firstoftwo
 \else \expandafter \@secondoftwo
 \fi
}%
\providecommand \@ifx [1]{%
 \ifx #1\expandafter \@firstoftwo
 \else \expandafter \@secondoftwo
 \fi
}%
\providecommand \natexlab [1]{#1}%
\providecommand \enquote  [1]{``#1''}%
\providecommand \bibnamefont  [1]{#1}%
\providecommand \bibfnamefont [1]{#1}%
\providecommand \citenamefont [1]{#1}%
\providecommand \href@noop [0]{\@secondoftwo}%
\providecommand \href [0]{\begingroup \@sanitize@url \@href}%
\providecommand \@href[1]{\@@startlink{#1}\@@href}%
\providecommand \@@href[1]{\endgroup#1\@@endlink}%
\providecommand \@sanitize@url [0]{\catcode `\\12\catcode `\$12\catcode
  `\&12\catcode `\#12\catcode `\^12\catcode `\_12\catcode `\%12\relax}%
\providecommand \@@startlink[1]{}%
\providecommand \@@endlink[0]{}%
\providecommand \url  [0]{\begingroup\@sanitize@url \@url }%
\providecommand \@url [1]{\endgroup\@href {#1}{\urlprefix }}%
\providecommand \urlprefix  [0]{URL }%
\providecommand \Eprint [0]{\href }%
\providecommand \doibase [0]{http://dx.doi.org/}%
\providecommand \selectlanguage [0]{\@gobble}%
\providecommand \bibinfo  [0]{\@secondoftwo}%
\providecommand \bibfield  [0]{\@secondoftwo}%
\providecommand \translation [1]{[#1]}%
\providecommand \BibitemOpen [0]{}%
\providecommand \bibitemStop [0]{}%
\providecommand \bibitemNoStop [0]{.\EOS\space}%
\providecommand \EOS [0]{\spacefactor3000\relax}%
\providecommand \BibitemShut  [1]{\csname bibitem#1\endcsname}%
\let\auto@bib@innerbib\@empty
\bibitem [{\citenamefont {Novoselov}\ \emph {et~al.}(2004)\citenamefont
  {Novoselov}, \citenamefont {Geim}, \citenamefont {Morozov}, \citenamefont
  {Jiang}, \citenamefont {Zhang}, \citenamefont {Dubonos}, \citenamefont
  {Grigorieva},\ and\ \citenamefont {Firsov}}]{Novoselov2004}%
  \BibitemOpen
  \bibfield  {author} {\bibinfo {author} {\bibfnamefont {K.~S.}\ \bibnamefont
  {Novoselov}}, \bibinfo {author} {\bibfnamefont {A.~K.}\ \bibnamefont {Geim}},
  \bibinfo {author} {\bibfnamefont {S.~V.}\ \bibnamefont {Morozov}}, \bibinfo
  {author} {\bibfnamefont {D.}~\bibnamefont {Jiang}}, \bibinfo {author}
  {\bibfnamefont {Y.}~\bibnamefont {Zhang}}, \bibinfo {author} {\bibfnamefont
  {S.~V.}\ \bibnamefont {Dubonos}}, \bibinfo {author} {\bibfnamefont {I.~V.}\
  \bibnamefont {Grigorieva}}, \ and\ \bibinfo {author} {\bibfnamefont {A.~A.}\
  \bibnamefont {Firsov}},\ }\href {\doibase 10.1126/science.1102896} {\bibfield
   {journal} {\bibinfo  {journal} {Science}\ }\textbf {\bibinfo {volume}
  {306}},\ \bibinfo {pages} {666} (\bibinfo {year} {2004})}\BibitemShut
  {NoStop}%
\bibitem [{\citenamefont {Meyer}\ \emph {et~al.}(2007)\citenamefont {Meyer},
  \citenamefont {Geim}, \citenamefont {Katsnelson}, \citenamefont {Novoselov},
  \citenamefont {Booth},\ and\ \citenamefont {Roth}}]{Meyer2007}%
  \BibitemOpen
  \bibfield  {author} {\bibinfo {author} {\bibfnamefont {J.~C.}\ \bibnamefont
  {Meyer}}, \bibinfo {author} {\bibfnamefont {A.~K.}\ \bibnamefont {Geim}},
  \bibinfo {author} {\bibfnamefont {M.~I.}\ \bibnamefont {Katsnelson}},
  \bibinfo {author} {\bibfnamefont {K.~S.}\ \bibnamefont {Novoselov}}, \bibinfo
  {author} {\bibfnamefont {T.~J.}\ \bibnamefont {Booth}}, \ and\ \bibinfo
  {author} {\bibfnamefont {S.}~\bibnamefont {Roth}},\ }\href {\doibase
  10.1038/nature05545} {\bibfield  {journal} {\bibinfo  {journal} {Nature
  (London)}\ }\textbf {\bibinfo {volume} {446}},\ \bibinfo {pages} {60}
  (\bibinfo {year} {2007})}\BibitemShut {NoStop}%
\bibitem [{\citenamefont {de~Heer}\ \emph {et~al.}(2007)\citenamefont
  {de~Heer}, \citenamefont {Berger}, \citenamefont {Wu}, \citenamefont {First},
  \citenamefont {Conrad}, \citenamefont {Li}, \citenamefont {Li}, \citenamefont
  {Sprinkle}, \citenamefont {Hass}, \citenamefont {Sadowski}, \citenamefont
  {Potemski},\ and\ \citenamefont {Martinez}}]{deHeer2007}%
  \BibitemOpen
  \bibfield  {author} {\bibinfo {author} {\bibfnamefont {W.~A.}\ \bibnamefont
  {de~Heer}}, \bibinfo {author} {\bibfnamefont {C.}~\bibnamefont {Berger}},
  \bibinfo {author} {\bibfnamefont {X.}~\bibnamefont {Wu}}, \bibinfo {author}
  {\bibfnamefont {P.~N.}\ \bibnamefont {First}}, \bibinfo {author}
  {\bibfnamefont {E.~H.}\ \bibnamefont {Conrad}}, \bibinfo {author}
  {\bibfnamefont {X.}~\bibnamefont {Li}}, \bibinfo {author} {\bibfnamefont
  {T.}~\bibnamefont {Li}}, \bibinfo {author} {\bibfnamefont {M.}~\bibnamefont
  {Sprinkle}}, \bibinfo {author} {\bibfnamefont {J.}~\bibnamefont {Hass}},
  \bibinfo {author} {\bibfnamefont {M.~L.}\ \bibnamefont {Sadowski}}, \bibinfo
  {author} {\bibfnamefont {M.}~\bibnamefont {Potemski}}, \ and\ \bibinfo
  {author} {\bibfnamefont {G.}~\bibnamefont {Martinez}},\ }\href {\doibase
  10.1016/j.ssc.2007.04.023} {\bibfield  {journal} {\bibinfo  {journal} {Solid
  State Commun.}\ }\textbf {\bibinfo {volume} {143}},\ \bibinfo {pages} {92 }
  (\bibinfo {year} {2007})}\BibitemShut {NoStop}%
\bibitem [{\citenamefont {Fasolino}, \citenamefont {Los},\ and\ \citenamefont
  {Katsnelson}(2007)}]{Fasolino2007}%
  \BibitemOpen
  \bibfield  {author} {\bibinfo {author} {\bibfnamefont {A.}~\bibnamefont
  {Fasolino}}, \bibinfo {author} {\bibfnamefont {J.~H.}\ \bibnamefont {Los}}, \
  and\ \bibinfo {author} {\bibfnamefont {M.~I.}\ \bibnamefont {Katsnelson}},\
  }\href {\doibase 10.1038/nmat2011} {\bibfield  {journal} {\bibinfo  {journal}
  {Nat. Mater.}\ }\textbf {\bibinfo {volume} {6}},\ \bibinfo {pages} {858}
  (\bibinfo {year} {2007})}\BibitemShut {NoStop}%
\bibitem [{\citenamefont {Guinea}, \citenamefont {Horovitz},\ and\
  \citenamefont {Le~Doussal}(2008)}]{Guinea2008may}%
  \BibitemOpen
  \bibfield  {author} {\bibinfo {author} {\bibfnamefont {F.}~\bibnamefont
  {Guinea}}, \bibinfo {author} {\bibfnamefont {B.}~\bibnamefont {Horovitz}}, \
  and\ \bibinfo {author} {\bibfnamefont {P.}~\bibnamefont {Le~Doussal}},\
  }\href {\doibase 10.1103/PhysRevB.77.205421} {\bibfield  {journal} {\bibinfo
  {journal} {Phys. Rev. B}\ }\textbf {\bibinfo {volume} {77}},\ \bibinfo
  {pages} {205421} (\bibinfo {year} {2008})}\BibitemShut {NoStop}%
\bibitem [{\citenamefont {Pereira}\ \emph {et~al.}(2010)\citenamefont
  {Pereira}, \citenamefont {Castro~Neto}, \citenamefont {Liang},\ and\
  \citenamefont {Mahadevan}}]{Pereira2010}%
  \BibitemOpen
  \bibfield  {author} {\bibinfo {author} {\bibfnamefont {V.~M.}\ \bibnamefont
  {Pereira}}, \bibinfo {author} {\bibfnamefont {A.~H.}\ \bibnamefont
  {Castro~Neto}}, \bibinfo {author} {\bibfnamefont {H.~Y.}\ \bibnamefont
  {Liang}}, \ and\ \bibinfo {author} {\bibfnamefont {L.}~\bibnamefont
  {Mahadevan}},\ }\href {\doibase 10.1103/PhysRevLett.105.156603} {\bibfield
  {journal} {\bibinfo  {journal} {Phys. Rev. Lett.}\ }\textbf {\bibinfo
  {volume} {105}},\ \bibinfo {pages} {156603} (\bibinfo {year}
  {2010})}\BibitemShut {NoStop}%
\bibitem [{\citenamefont {Levy}\ \emph {et~al.}(2010)\citenamefont {Levy},
  \citenamefont {Burke}, \citenamefont {Meaker}, \citenamefont {Panlasigui},
  \citenamefont {Zettl}, \citenamefont {Guinea}, \citenamefont {Neto},\ and\
  \citenamefont {Crommie}}]{Levy2010}%
  \BibitemOpen
  \bibfield  {author} {\bibinfo {author} {\bibfnamefont {N.}~\bibnamefont
  {Levy}}, \bibinfo {author} {\bibfnamefont {S.~A.}\ \bibnamefont {Burke}},
  \bibinfo {author} {\bibfnamefont {K.~L.}\ \bibnamefont {Meaker}}, \bibinfo
  {author} {\bibfnamefont {M.}~\bibnamefont {Panlasigui}}, \bibinfo {author}
  {\bibfnamefont {A.}~\bibnamefont {Zettl}}, \bibinfo {author} {\bibfnamefont
  {F.}~\bibnamefont {Guinea}}, \bibinfo {author} {\bibfnamefont {A.~H.~C.}\
  \bibnamefont {Neto}}, \ and\ \bibinfo {author} {\bibfnamefont {M.~F.}\
  \bibnamefont {Crommie}},\ }\href {\doibase 10.1126/science.1191700}
  {\bibfield  {journal} {\bibinfo  {journal} {Science}\ }\textbf {\bibinfo
  {volume} {329}},\ \bibinfo {pages} {544} (\bibinfo {year}
  {2010})}\BibitemShut {NoStop}%
\bibitem [{\citenamefont {Balog}\ \emph {et~al.}(2010)\citenamefont {Balog},
  \citenamefont {Jorgensen}, \citenamefont {Nilsson}, \citenamefont {Andersen},
  \citenamefont {Rienks}, \citenamefont {Bianchi}, \citenamefont {Fanetti},
  \citenamefont {Laegsgaard}, \citenamefont {Baraldi}, \citenamefont {Lizzit},
  \citenamefont {Sljivancanin}, \citenamefont {Besenbacher}, \citenamefont
  {Hammer}, \citenamefont {Pedersen}, \citenamefont {Hofmann},\ and\
  \citenamefont {Hornekaer}}]{Balog2010}%
  \BibitemOpen
  \bibfield  {author} {\bibinfo {author} {\bibfnamefont {R.}~\bibnamefont
  {Balog}}, \bibinfo {author} {\bibfnamefont {B.}~\bibnamefont {Jorgensen}},
  \bibinfo {author} {\bibfnamefont {L.}~\bibnamefont {Nilsson}}, \bibinfo
  {author} {\bibfnamefont {M.}~\bibnamefont {Andersen}}, \bibinfo {author}
  {\bibfnamefont {E.}~\bibnamefont {Rienks}}, \bibinfo {author} {\bibfnamefont
  {M.}~\bibnamefont {Bianchi}}, \bibinfo {author} {\bibfnamefont
  {M.}~\bibnamefont {Fanetti}}, \bibinfo {author} {\bibfnamefont
  {E.}~\bibnamefont {Laegsgaard}}, \bibinfo {author} {\bibfnamefont
  {A.}~\bibnamefont {Baraldi}}, \bibinfo {author} {\bibfnamefont
  {S.}~\bibnamefont {Lizzit}}, \bibinfo {author} {\bibfnamefont
  {Z.}~\bibnamefont {Sljivancanin}}, \bibinfo {author} {\bibfnamefont
  {F.}~\bibnamefont {Besenbacher}}, \bibinfo {author} {\bibfnamefont
  {B.}~\bibnamefont {Hammer}}, \bibinfo {author} {\bibfnamefont {T.~G.}\
  \bibnamefont {Pedersen}}, \bibinfo {author} {\bibfnamefont {P.}~\bibnamefont
  {Hofmann}}, \ and\ \bibinfo {author} {\bibfnamefont {L.}~\bibnamefont
  {Hornekaer}},\ }\href {\doibase 10.1038/NMAT2710} {\bibfield  {journal}
  {\bibinfo  {journal} {Nat. Mater.}\ }\textbf {\bibinfo {volume} {9}},\
  \bibinfo {pages} {315} (\bibinfo {year} {2010})}\BibitemShut {NoStop}%
\bibitem [{\citenamefont {Georgiou}\ \emph {et~al.}(2011)\citenamefont
  {Georgiou}, \citenamefont {Britnell}, \citenamefont {Blake}, \citenamefont
  {Gorbachev}, \citenamefont {Gholinia}, \citenamefont {Geim}, \citenamefont
  {Casiraghi},\ and\ \citenamefont {Novoselov}}]{Georgiou2011}%
  \BibitemOpen
  \bibfield  {author} {\bibinfo {author} {\bibfnamefont {T.}~\bibnamefont
  {Georgiou}}, \bibinfo {author} {\bibfnamefont {L.}~\bibnamefont {Britnell}},
  \bibinfo {author} {\bibfnamefont {P.}~\bibnamefont {Blake}}, \bibinfo
  {author} {\bibfnamefont {R.~V.}\ \bibnamefont {Gorbachev}}, \bibinfo {author}
  {\bibfnamefont {A.}~\bibnamefont {Gholinia}}, \bibinfo {author}
  {\bibfnamefont {A.~K.}\ \bibnamefont {Geim}}, \bibinfo {author}
  {\bibfnamefont {C.}~\bibnamefont {Casiraghi}}, \ and\ \bibinfo {author}
  {\bibfnamefont {K.~S.}\ \bibnamefont {Novoselov}},\ }\href {\doibase
  10.1063/1.3631632} {\bibfield  {journal} {\bibinfo  {journal} {Appl. Phys.
  Lett.}\ }\textbf {\bibinfo {volume} {99}},\ \bibinfo {eid} {093103} (\bibinfo
  {year} {2011})}\BibitemShut {NoStop}%
\bibitem [{\citenamefont {Wang}, \citenamefont {Mylvaganam},\ and\
  \citenamefont {Zhang}(2009)}]{Wang2009}%
  \BibitemOpen
  \bibfield  {author} {\bibinfo {author} {\bibfnamefont {C.~Y.}\ \bibnamefont
  {Wang}}, \bibinfo {author} {\bibfnamefont {K.}~\bibnamefont {Mylvaganam}}, \
  and\ \bibinfo {author} {\bibfnamefont {L.~C.}\ \bibnamefont {Zhang}},\ }\href
  {\doibase 10.1103/PhysRevB.80.155445} {\bibfield  {journal} {\bibinfo
  {journal} {Phys. Rev. B}\ }\textbf {\bibinfo {volume} {80}},\ \bibinfo
  {pages} {155445} (\bibinfo {year} {2009})}\BibitemShut {NoStop}%
\bibitem [{\citenamefont {Gil}\ \emph {et~al.}(2010)\citenamefont {Gil},
  \citenamefont {Adhikari}, \citenamefont {Scarpa},\ and\ \citenamefont
  {Bonet}}]{Gil2010}%
  \BibitemOpen
  \bibfield  {author} {\bibinfo {author} {\bibfnamefont {A.~J.}\ \bibnamefont
  {Gil}}, \bibinfo {author} {\bibfnamefont {S.}~\bibnamefont {Adhikari}},
  \bibinfo {author} {\bibfnamefont {F.}~\bibnamefont {Scarpa}}, \ and\ \bibinfo
  {author} {\bibfnamefont {J.}~\bibnamefont {Bonet}},\ }\href {\doibase
  10.1088/0953-8984/22/14/145302} {\bibfield  {journal} {\bibinfo  {journal}
  {J. Phys.: Condens. Matter}\ }\textbf {\bibinfo {volume} {22}},\ \bibinfo
  {pages} {145302} (\bibinfo {year} {2010})}\BibitemShut {NoStop}%
\bibitem [{\citenamefont {Duan}, \citenamefont {Gong},\ and\ \citenamefont
  {Wang}(2011)}]{Duan2011}%
  \BibitemOpen
  \bibfield  {author} {\bibinfo {author} {\bibfnamefont {W.~H.}\ \bibnamefont
  {Duan}}, \bibinfo {author} {\bibfnamefont {K.}~\bibnamefont {Gong}}, \ and\
  \bibinfo {author} {\bibfnamefont {Q.}~\bibnamefont {Wang}},\ }\href {\doibase
  10.1016/j.carbon.2011.03.033} {\bibfield  {journal} {\bibinfo  {journal}
  {Carbon}\ }\textbf {\bibinfo {volume} {49}},\ \bibinfo {pages} {3107 }
  (\bibinfo {year} {2011})}\BibitemShut {NoStop}%
\bibitem [{\citenamefont {Min}\ and\ \citenamefont {Aluru}(2011)}]{Min2011}%
  \BibitemOpen
  \bibfield  {author} {\bibinfo {author} {\bibfnamefont {K.}~\bibnamefont
  {Min}}\ and\ \bibinfo {author} {\bibfnamefont {N.~R.}\ \bibnamefont
  {Aluru}},\ }\href {\doibase 10.1063/1.3534787} {\bibfield  {journal}
  {\bibinfo  {journal} {Appl. Phys. Lett.}\ }\textbf {\bibinfo {volume} {98}},\
  \bibinfo {eid} {013113} (\bibinfo {year} {2011})}\BibitemShut {NoStop}%
\bibitem [{\citenamefont {Liu}\ \emph {et~al.}(2011)\citenamefont {Liu},
  \citenamefont {Pan}, \citenamefont {Fu}, \citenamefont {Zhang}, \citenamefont
  {Dai},\ and\ \citenamefont {Liu}}]{Liu2011}%
  \BibitemOpen
  \bibfield  {author} {\bibinfo {author} {\bibfnamefont {N.}~\bibnamefont
  {Liu}}, \bibinfo {author} {\bibfnamefont {Z.}~\bibnamefont {Pan}}, \bibinfo
  {author} {\bibfnamefont {L.}~\bibnamefont {Fu}}, \bibinfo {author}
  {\bibfnamefont {C.}~\bibnamefont {Zhang}}, \bibinfo {author} {\bibfnamefont
  {B.}~\bibnamefont {Dai}}, \ and\ \bibinfo {author} {\bibfnamefont
  {Z.}~\bibnamefont {Liu}},\ }\href {\doibase 10.1007/s12274-011-0156-3}
  {\bibfield  {journal} {\bibinfo  {journal} {Nano Res.}\ }\textbf {\bibinfo
  {volume} {4}},\ \bibinfo {pages} {996} (\bibinfo {year} {2011})}\BibitemShut
  {NoStop}%
\bibitem [{\citenamefont {Ni}\ \emph {et~al.}(2012)\citenamefont {Ni},
  \citenamefont {Zheng}, \citenamefont {Bae}, \citenamefont {Kim},
  \citenamefont {Pachoud}, \citenamefont {Kim}, \citenamefont {Tan},
  \citenamefont {Im}, \citenamefont {Ahn}, \citenamefont {Hong},\ and\
  \citenamefont {Özyilmaz}}]{Ni2012}%
  \BibitemOpen
  \bibfield  {author} {\bibinfo {author} {\bibfnamefont {G.-X.}\ \bibnamefont
  {Ni}}, \bibinfo {author} {\bibfnamefont {Y.}~\bibnamefont {Zheng}}, \bibinfo
  {author} {\bibfnamefont {S.}~\bibnamefont {Bae}}, \bibinfo {author}
  {\bibfnamefont {H.~R.}\ \bibnamefont {Kim}}, \bibinfo {author} {\bibfnamefont
  {A.}~\bibnamefont {Pachoud}}, \bibinfo {author} {\bibfnamefont {Y.~S.}\
  \bibnamefont {Kim}}, \bibinfo {author} {\bibfnamefont {C.-L.}\ \bibnamefont
  {Tan}}, \bibinfo {author} {\bibfnamefont {D.}~\bibnamefont {Im}}, \bibinfo
  {author} {\bibfnamefont {J.-H.}\ \bibnamefont {Ahn}}, \bibinfo {author}
  {\bibfnamefont {B.~H.}\ \bibnamefont {Hong}}, \ and\ \bibinfo {author}
  {\bibfnamefont {B.}~\bibnamefont {Özyilmaz}},\ }\href {\doibase
  10.1021/nn203775x} {\bibfield  {journal} {\bibinfo  {journal} {ACS Nano}\
  }\textbf {\bibinfo {volume} {6}},\ \bibinfo {pages} {1158} (\bibinfo {year}
  {2012})}\BibitemShut {NoStop}%
\bibitem [{\citenamefont {Gao}\ \emph {et~al.}(2012)\citenamefont {Gao},
  \citenamefont {Ren}, \citenamefont {Xu}, \citenamefont {Jin}, \citenamefont
  {Wang}, \citenamefont {Ma}, \citenamefont {Ma}, \citenamefont {Zhang},
  \citenamefont {Fu}, \citenamefont {Peng}, \citenamefont {Bao},\ and\
  \citenamefont {Cheng}}]{Gao2012}%
  \BibitemOpen
  \bibfield  {author} {\bibinfo {author} {\bibfnamefont {L.}~\bibnamefont
  {Gao}}, \bibinfo {author} {\bibfnamefont {W.}~\bibnamefont {Ren}}, \bibinfo
  {author} {\bibfnamefont {H.}~\bibnamefont {Xu}}, \bibinfo {author}
  {\bibfnamefont {L.}~\bibnamefont {Jin}}, \bibinfo {author} {\bibfnamefont
  {Z.}~\bibnamefont {Wang}}, \bibinfo {author} {\bibfnamefont {T.}~\bibnamefont
  {Ma}}, \bibinfo {author} {\bibfnamefont {L.-P.}\ \bibnamefont {Ma}}, \bibinfo
  {author} {\bibfnamefont {Z.}~\bibnamefont {Zhang}}, \bibinfo {author}
  {\bibfnamefont {Q.}~\bibnamefont {Fu}}, \bibinfo {author} {\bibfnamefont
  {L.-M.}\ \bibnamefont {Peng}}, \bibinfo {author} {\bibfnamefont
  {X.}~\bibnamefont {Bao}}, \ and\ \bibinfo {author} {\bibfnamefont {H.-M.}\
  \bibnamefont {Cheng}},\ }\href {\doibase 10.1038/ncomms1702} {\bibfield
  {journal} {\bibinfo  {journal} {Nat. Commun.}\ }\textbf {\bibinfo {volume}
  {3}},\ \bibinfo {pages} {699} (\bibinfo {year} {2012})}\BibitemShut {NoStop}%
\bibitem [{\citenamefont {Zhu}\ \emph {et~al.}(2012)\citenamefont {Zhu},
  \citenamefont {Low}, \citenamefont {Perebeinos}, \citenamefont {Bol},
  \citenamefont {Zhu}, \citenamefont {Yan}, \citenamefont {Tersoff},\ and\
  \citenamefont {Avouris}}]{Zhu2012}%
  \BibitemOpen
  \bibfield  {author} {\bibinfo {author} {\bibfnamefont {W.}~\bibnamefont
  {Zhu}}, \bibinfo {author} {\bibfnamefont {T.}~\bibnamefont {Low}}, \bibinfo
  {author} {\bibfnamefont {V.}~\bibnamefont {Perebeinos}}, \bibinfo {author}
  {\bibfnamefont {A.~A.}\ \bibnamefont {Bol}}, \bibinfo {author} {\bibfnamefont
  {Y.}~\bibnamefont {Zhu}}, \bibinfo {author} {\bibfnamefont {H.}~\bibnamefont
  {Yan}}, \bibinfo {author} {\bibfnamefont {J.}~\bibnamefont {Tersoff}}, \ and\
  \bibinfo {author} {\bibfnamefont {P.}~\bibnamefont {Avouris}},\ }\href
  {\doibase 10.1021/nl300563h} {\bibfield  {journal} {\bibinfo  {journal} {Nano
  Letters}\ }\textbf {\bibinfo {volume} {ASAP}} (\bibinfo {year} {2012}),\
  10.1021/nl300563h}\BibitemShut {NoStop}%
\bibitem [{\citenamefont {Ishigami}\ \emph {et~al.}(2007)\citenamefont
  {Ishigami}, \citenamefont {Chen}, \citenamefont {Cullen}, \citenamefont
  {Fuhrer},\ and\ \citenamefont {Williams}}]{Ishigami2007}%
  \BibitemOpen
  \bibfield  {author} {\bibinfo {author} {\bibfnamefont {M.}~\bibnamefont
  {Ishigami}}, \bibinfo {author} {\bibfnamefont {J.~H.}\ \bibnamefont {Chen}},
  \bibinfo {author} {\bibfnamefont {W.~G.}\ \bibnamefont {Cullen}}, \bibinfo
  {author} {\bibfnamefont {M.~S.}\ \bibnamefont {Fuhrer}}, \ and\ \bibinfo
  {author} {\bibfnamefont {E.~D.}\ \bibnamefont {Williams}},\ }\href {\doibase
  10.1021/nl070613a} {\bibfield  {journal} {\bibinfo  {journal} {Nano Lett.}\
  }\textbf {\bibinfo {volume} {7}},\ \bibinfo {pages} {1643} (\bibinfo {year}
  {2007})}\BibitemShut {NoStop}%
\bibitem [{\citenamefont {Xu}, \citenamefont {Cao},\ and\ \citenamefont
  {Heath}(2009)}]{Xu2009}%
  \BibitemOpen
  \bibfield  {author} {\bibinfo {author} {\bibfnamefont {K.}~\bibnamefont
  {Xu}}, \bibinfo {author} {\bibfnamefont {P.}~\bibnamefont {Cao}}, \ and\
  \bibinfo {author} {\bibfnamefont {J.~R.}\ \bibnamefont {Heath}},\ }\href
  {\doibase 10.1021/nl902729p} {\bibfield  {journal} {\bibinfo  {journal} {Nano
  Lett.}\ }\textbf {\bibinfo {volume} {9}},\ \bibinfo {pages} {4446} (\bibinfo
  {year} {2009})}\BibitemShut {NoStop}%
\bibitem [{\citenamefont {Schneider}\ \emph {et~al.}(2010)\citenamefont
  {Schneider}, \citenamefont {Calado}, \citenamefont {Zandbergen},
  \citenamefont {Vandersypen},\ and\ \citenamefont {Dekker}}]{Schneider2010a}%
  \BibitemOpen
  \bibfield  {author} {\bibinfo {author} {\bibfnamefont {G.~F.}\ \bibnamefont
  {Schneider}}, \bibinfo {author} {\bibfnamefont {V.~E.}\ \bibnamefont
  {Calado}}, \bibinfo {author} {\bibfnamefont {H.}~\bibnamefont {Zandbergen}},
  \bibinfo {author} {\bibfnamefont {L.~M.~K.}\ \bibnamefont {Vandersypen}}, \
  and\ \bibinfo {author} {\bibfnamefont {C.}~\bibnamefont {Dekker}},\ }\href
  {\doibase 10.1021/nl1008037} {\bibfield  {journal} {\bibinfo  {journal} {Nano
  Lett.}\ }\textbf {\bibinfo {volume} {10}},\ \bibinfo {pages} {1912} (\bibinfo
  {year} {2010})}\BibitemShut {NoStop}%
\bibitem [{\citenamefont {Ferrari}\ \emph {et~al.}(2006)\citenamefont
  {Ferrari}, \citenamefont {Meyer}, \citenamefont {Scardaci}, \citenamefont
  {Casiraghi}, \citenamefont {Lazzeri}, \citenamefont {Mauri}, \citenamefont
  {Piscanec}, \citenamefont {Jiang}, \citenamefont {Novoselov}, \citenamefont
  {Roth},\ and\ \citenamefont {Geim}}]{Ferrari2006}%
  \BibitemOpen
  \bibfield  {author} {\bibinfo {author} {\bibfnamefont {A.~C.}\ \bibnamefont
  {Ferrari}}, \bibinfo {author} {\bibfnamefont {J.~C.}\ \bibnamefont {Meyer}},
  \bibinfo {author} {\bibfnamefont {V.}~\bibnamefont {Scardaci}}, \bibinfo
  {author} {\bibfnamefont {C.}~\bibnamefont {Casiraghi}}, \bibinfo {author}
  {\bibfnamefont {M.}~\bibnamefont {Lazzeri}}, \bibinfo {author} {\bibfnamefont
  {F.}~\bibnamefont {Mauri}}, \bibinfo {author} {\bibfnamefont
  {S.}~\bibnamefont {Piscanec}}, \bibinfo {author} {\bibfnamefont
  {D.}~\bibnamefont {Jiang}}, \bibinfo {author} {\bibfnamefont {K.~S.}\
  \bibnamefont {Novoselov}}, \bibinfo {author} {\bibfnamefont {S.}~\bibnamefont
  {Roth}}, \ and\ \bibinfo {author} {\bibfnamefont {A.~K.}\ \bibnamefont
  {Geim}},\ }\href {\doibase 10.1103/PhysRevLett.97.187401} {\bibfield
  {journal} {\bibinfo  {journal} {Phys. Rev. Lett.}\ }\textbf {\bibinfo
  {volume} {97}},\ \bibinfo {pages} {187401} (\bibinfo {year}
  {2006})}\BibitemShut {NoStop}%
\bibitem [{\citenamefont {Guinea}, \citenamefont {Katsnelson},\ and\
  \citenamefont {Vozmediano}(2008)}]{Guinea2008feb}%
  \BibitemOpen
  \bibfield  {author} {\bibinfo {author} {\bibfnamefont {F.}~\bibnamefont
  {Guinea}}, \bibinfo {author} {\bibfnamefont {M.~I.}\ \bibnamefont
  {Katsnelson}}, \ and\ \bibinfo {author} {\bibfnamefont {M.~A.~H.}\
  \bibnamefont {Vozmediano}},\ }\href {\doibase 10.1103/PhysRevB.77.075422}
  {\bibfield  {journal} {\bibinfo  {journal} {Phys. Rev. B}\ }\textbf {\bibinfo
  {volume} {77}},\ \bibinfo {pages} {075422} (\bibinfo {year}
  {2008})}\BibitemShut {NoStop}%
\bibitem [{\citenamefont {Elias}\ \emph {et~al.}(2009)\citenamefont {Elias},
  \citenamefont {Nair}, \citenamefont {Mohiuddin}, \citenamefont {Morozov},
  \citenamefont {Blake}, \citenamefont {Halsall}, \citenamefont {Ferrari},
  \citenamefont {Boukhvalov}, \citenamefont {Katsnelson}, \citenamefont
  {Geim},\ and\ \citenamefont {Novoselov}}]{Elias2009}%
  \BibitemOpen
  \bibfield  {author} {\bibinfo {author} {\bibfnamefont {D.~C.}\ \bibnamefont
  {Elias}}, \bibinfo {author} {\bibfnamefont {R.~R.}\ \bibnamefont {Nair}},
  \bibinfo {author} {\bibfnamefont {T.~M.~G.}\ \bibnamefont {Mohiuddin}},
  \bibinfo {author} {\bibfnamefont {S.~V.}\ \bibnamefont {Morozov}}, \bibinfo
  {author} {\bibfnamefont {P.}~\bibnamefont {Blake}}, \bibinfo {author}
  {\bibfnamefont {M.~P.}\ \bibnamefont {Halsall}}, \bibinfo {author}
  {\bibfnamefont {A.~C.}\ \bibnamefont {Ferrari}}, \bibinfo {author}
  {\bibfnamefont {D.~W.}\ \bibnamefont {Boukhvalov}}, \bibinfo {author}
  {\bibfnamefont {M.~I.}\ \bibnamefont {Katsnelson}}, \bibinfo {author}
  {\bibfnamefont {A.~K.}\ \bibnamefont {Geim}}, \ and\ \bibinfo {author}
  {\bibfnamefont {K.~S.}\ \bibnamefont {Novoselov}},\ }\href {\doibase
  10.1126/science.1167130} {\bibfield  {journal} {\bibinfo  {journal}
  {Science}\ }\textbf {\bibinfo {volume} {323}},\ \bibinfo {pages} {610}
  (\bibinfo {year} {2009})}\BibitemShut {NoStop}%
\bibitem [{\citenamefont {Nair}\ \emph {et~al.}(2010)\citenamefont {Nair},
  \citenamefont {Ren}, \citenamefont {Jalil}, \citenamefont {Riaz},
  \citenamefont {Kravets}, \citenamefont {Britnell}, \citenamefont {Blake},
  \citenamefont {Schedin}, \citenamefont {Mayorov}, \citenamefont {Yuan},
  \citenamefont {Katsnelson}, \citenamefont {Cheng}, \citenamefont
  {Strupinski}, \citenamefont {Bulusheva}, \citenamefont {Okotrub},
  \citenamefont {Grigorieva}, \citenamefont {Grigorenko}, \citenamefont
  {Novoselov},\ and\ \citenamefont {Geim}}]{Nair2010}%
  \BibitemOpen
  \bibfield  {author} {\bibinfo {author} {\bibfnamefont {R.~R.}\ \bibnamefont
  {Nair}}, \bibinfo {author} {\bibfnamefont {W.}~\bibnamefont {Ren}}, \bibinfo
  {author} {\bibfnamefont {R.}~\bibnamefont {Jalil}}, \bibinfo {author}
  {\bibfnamefont {I.}~\bibnamefont {Riaz}}, \bibinfo {author} {\bibfnamefont
  {V.~G.}\ \bibnamefont {Kravets}}, \bibinfo {author} {\bibfnamefont
  {L.}~\bibnamefont {Britnell}}, \bibinfo {author} {\bibfnamefont
  {P.}~\bibnamefont {Blake}}, \bibinfo {author} {\bibfnamefont
  {F.}~\bibnamefont {Schedin}}, \bibinfo {author} {\bibfnamefont {A.~S.}\
  \bibnamefont {Mayorov}}, \bibinfo {author} {\bibfnamefont {S.}~\bibnamefont
  {Yuan}}, \bibinfo {author} {\bibfnamefont {M.~I.}\ \bibnamefont
  {Katsnelson}}, \bibinfo {author} {\bibfnamefont {H.-M.}\ \bibnamefont
  {Cheng}}, \bibinfo {author} {\bibfnamefont {W.}~\bibnamefont {Strupinski}},
  \bibinfo {author} {\bibfnamefont {L.~G.}\ \bibnamefont {Bulusheva}}, \bibinfo
  {author} {\bibfnamefont {A.~V.}\ \bibnamefont {Okotrub}}, \bibinfo {author}
  {\bibfnamefont {I.~V.}\ \bibnamefont {Grigorieva}}, \bibinfo {author}
  {\bibfnamefont {A.~N.}\ \bibnamefont {Grigorenko}}, \bibinfo {author}
  {\bibfnamefont {K.~S.}\ \bibnamefont {Novoselov}}, \ and\ \bibinfo {author}
  {\bibfnamefont {A.~K.}\ \bibnamefont {Geim}},\ }\href {\doibase
  10.1002/smll.201090086} {\bibfield  {journal} {\bibinfo  {journal} {Small}\
  }\textbf {\bibinfo {volume} {6}},\ \bibinfo {pages} {2773} (\bibinfo {year}
  {2010})}\BibitemShut {NoStop}%
\end{thebibliography}%

\end{document}